\documentclass[11pt,twoside]{article}
\usepackage{asp2010}

\resetcounters

\begin{document}

\title{Extragalactic Nova Populations}
\author{A. W. Shafter,$^1$ C. Curtin,$^1$ C. J. Pritchet,$^2$ M. F. Bode,$^3$ and M. J. Darnley$^3$
\affil{$^1$Department of Astronomy, San Diego State University, San Diego, CA 92182, USA}
\affil{$^2$Department of Physics and Astronomy, University of Victoria, Victoria, BC, Canada}
\affil{$^3$Astrophysics Research Institute, Liverpool John Moores University, Birkenhead, CH41 1LD, UK}}

\begin{abstract}
Nova rates have now been measured for more than a dozen galaxies
spanning a wide range of Hubble types. When normalized to the
infrared $K$-band luminosity of the galaxy, the luminosity-specific
nova rates typically fall in the range of 1--3 novae per year per
$10^{10}$ solar luminosities in $K$, and do
not vary significantly across the Hubble sequence.
Preliminary nova rates are presented for three
Virgo ellipticals (M49, M84, and M87) with differing globular
cluster specific frequencies. No dependence of the luminosity-specific
nova rate on globular cluster specific frequency was found.
Photometric and spectroscopic observations of novae in the Local
Group suggest that galaxies dominated by
a younger stellar population (M33 and the LMC) are characterized by
novae with a generally faster photometric evolution, and by a higher
fraction of He/N novae compared with novae in M31. The recurrent nova
population in the LMC appears to be higher than that seen in M31 and
the Galaxy.

\end{abstract}

\section{Extragalactic Nova Rates}
Classical novae can reach absolute visual magnitudes of $V=-10$ making them
among the brightest transient sources known. Their high
luminosities, coupled with their frequency of appearance [$\sim35$~yr$^{-1}$
in a galaxy like our own \citep{sha97}]
make novae ideal for probing the properties
of close binary stars in different (extragalactic) stellar populations.
Theoretical models show that
the strength of the nova outburst is most sensitive to the mass of the
accreting white dwarf, with outbursts occurring on massive stars
expected to be brighter, with shorter recurrence times and faster light
curve evolution. Since population synthesis studies predict that the
mean white dwarf mass in a nova system will decrease as a function of the time
elapsed since the formation of the progenitor
binary \citep[e.g.,][]{tut95,pol96},
the proportion of fast and bright novae is
expected to be higher in younger stellar populations.

\begin{figure}
\includegraphics[clip=true, scale=0.50, angle=-90]{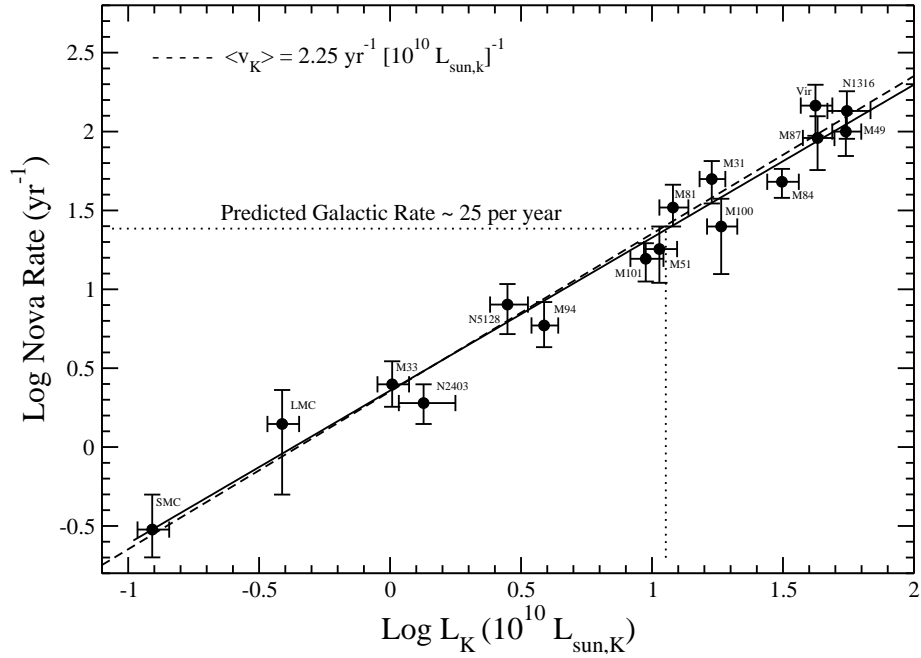}
\caption{The dependence of a galaxy's nova rate on its $K$-band luminosity. A Galactic nova rate of 25~yr$^{-1}$ is predicted based on the extragalactic scaling.}
\end{figure}
\begin{figure}
\includegraphics[clip=true, scale=0.47, angle=-90]{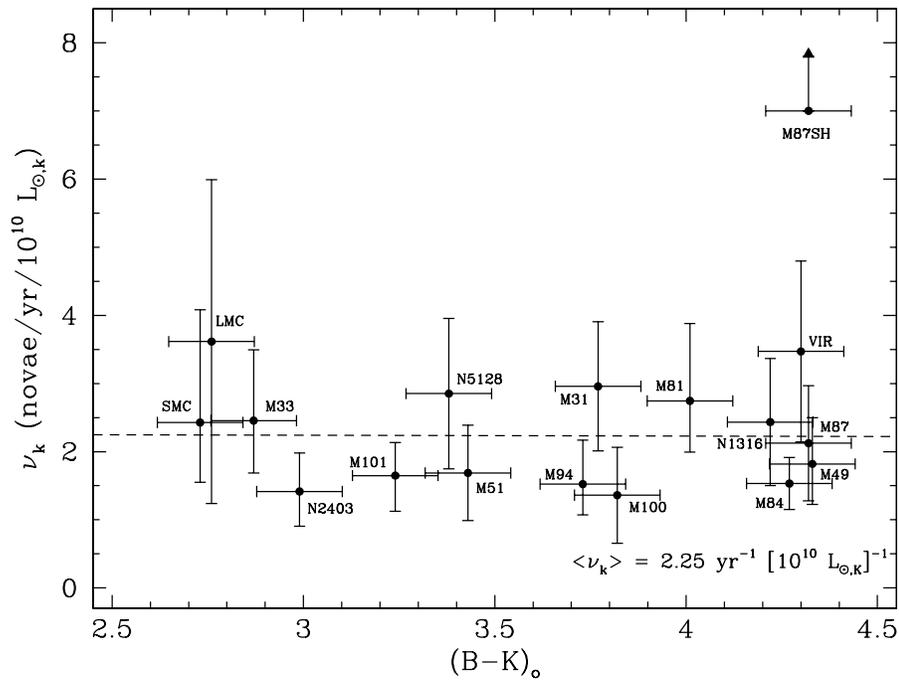}
\caption{The dependence of $K$-band LSNR on galaxy color. M87SH is based on
\citet{sha02}.}
\end{figure}

To date, novae have been observed in more than a dozen galaxies, some as distant
as the Coma Cluster, with
a sufficient sample of novae having been observed to estimate rates
in 15 galaxies. Figure~1 shows the nova rates plotted
as a function of the integrated $K$-band luminosity of the host galaxy.
The slope
of the relation in this log-log plot is consistent with unity, indicating that
the nova rate is simply proportional to the $K$-band luminosity of the galaxy.
The dashed line shows the best-fitting linear relation of the form
$R = \nu_K~L_K$, where $\nu_K = 2.25$~yr$^{-1}~[10^{10}L_{\odot,K}]^{-1}$.
The predicted Galactic rate of $\sim$25~yr$^{-1}$ is nearly 30\% lower
than the best direct estimate, raising the possibility
that the extragalactic nova rates may be systematically underestimated.
Figure~2 shows the luminosity-specific nova rates (LSNRs)
plotted as a function of
galaxy color. With the possible exception of M87 \citep{sha02, miz13},
there is no evidence that the LSNR varies significantly with Hubble type. 

\begin{figure}
\includegraphics[clip=true, scale=0.32, angle=-90]{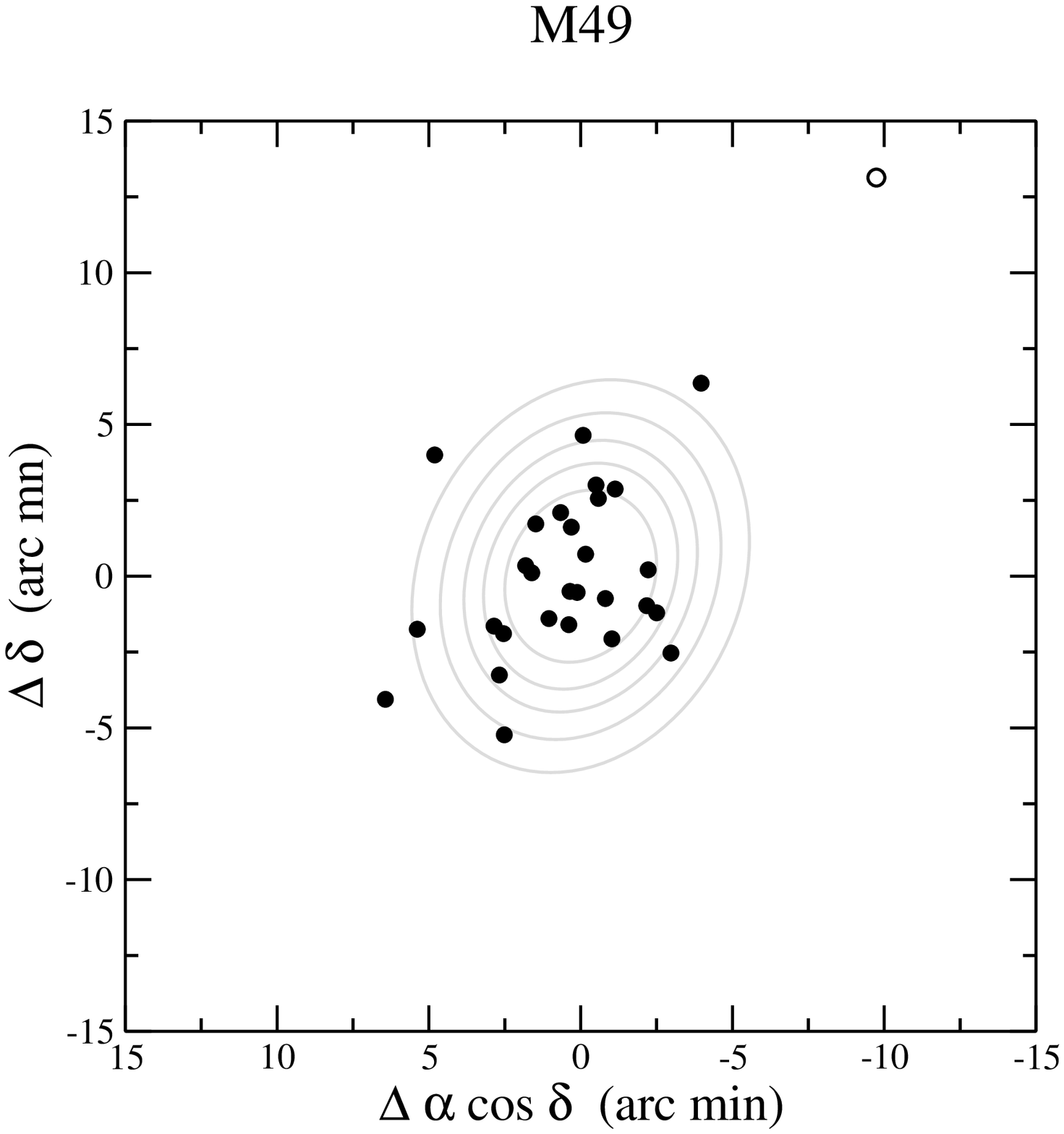}
\includegraphics[clip=true, scale=0.32, angle=-90]{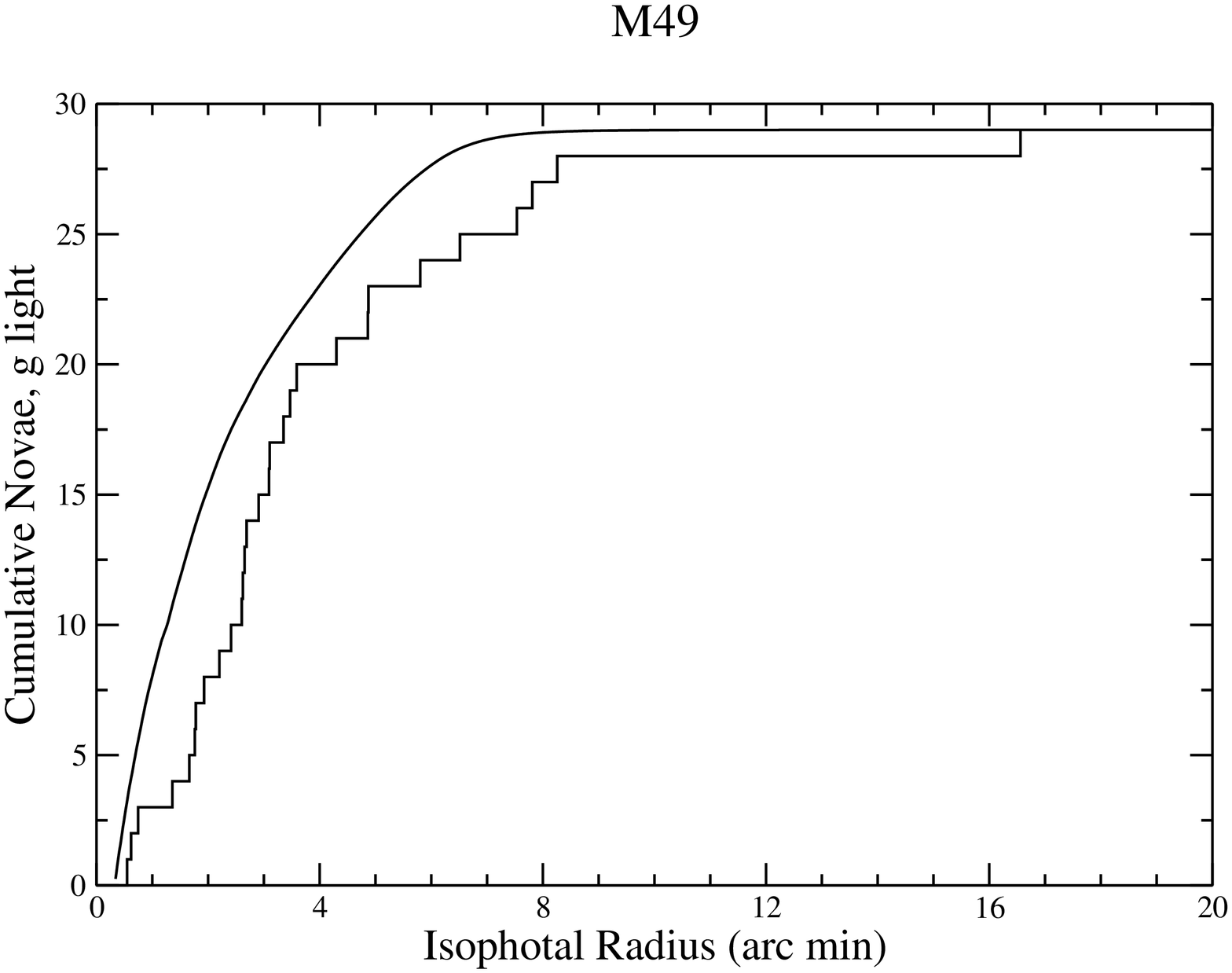}
\end{figure}
\begin{figure}
\includegraphics[clip=true, scale=0.32, angle=-90]{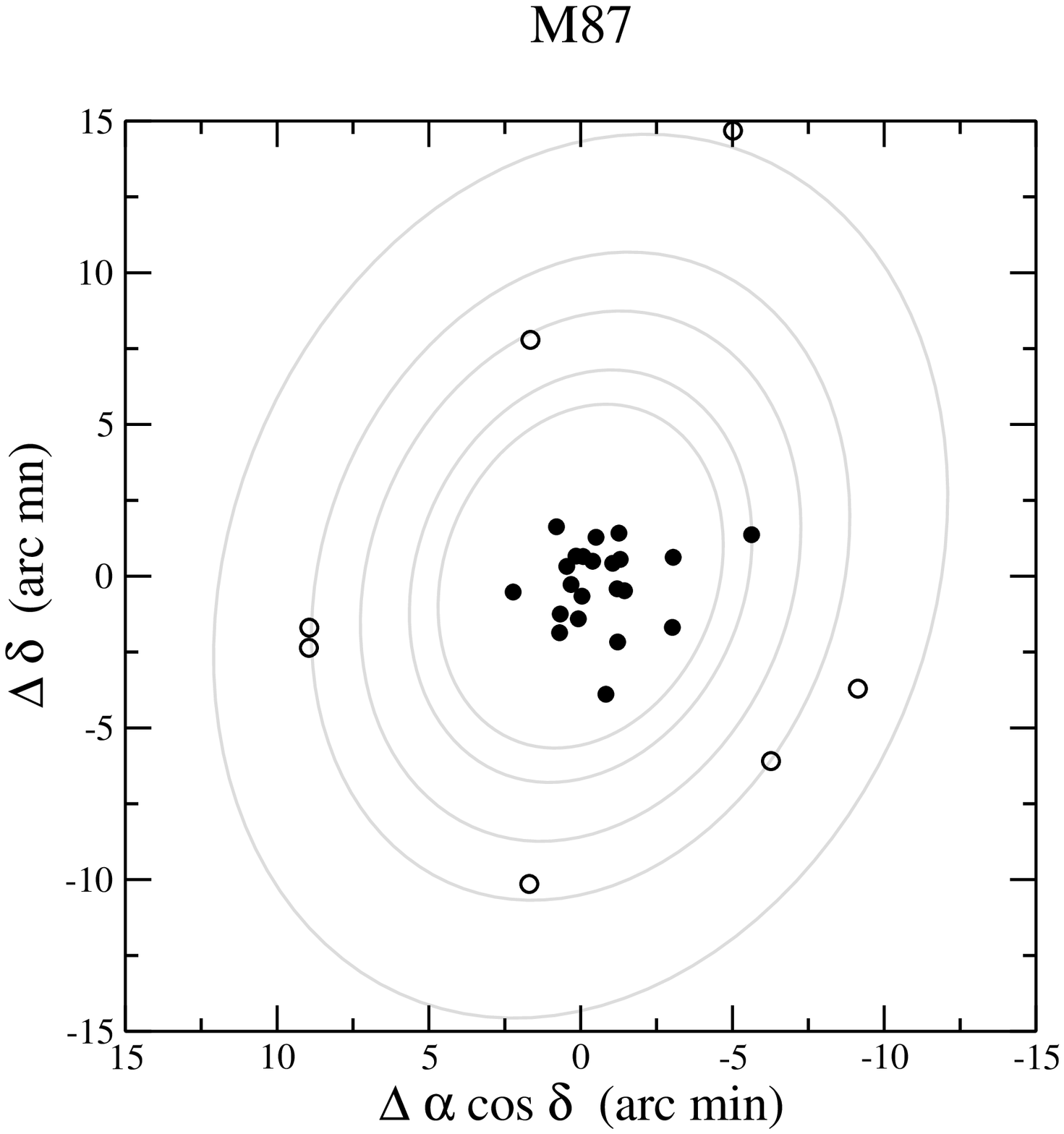}
\includegraphics[clip=true, scale=0.32, angle=-90]{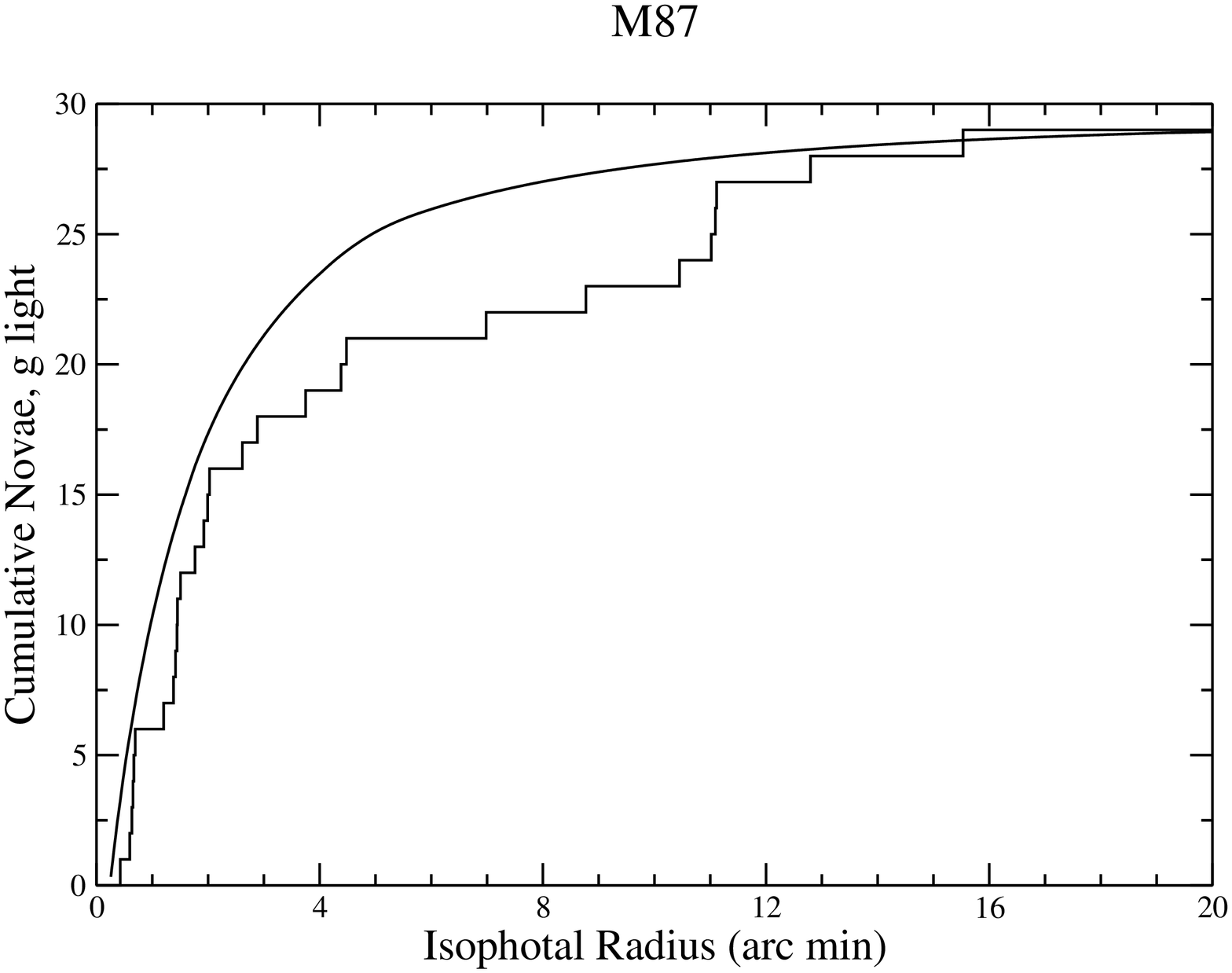}
\caption{The spatial distributions of novae in the Virgo
ellipticals M49 and M87 are shown along with a comparison of the
cumulative nova distributions and the background $g$-band light. Open
circles represent transients that may not be novae associated with the galaxies.}
\end{figure}

\subsection{Virgo Ellipticals}

The surprising result that novae in M31 appeared to be 
primarily associated with the galaxy's bulge population led
\citet{cia87} to propose that a significant fraction of
novae might be spawned in the galaxy's globular clusters and subsequently
ejected into the bulge through 3-body interactions in the clusters, or
through cluster disruption, or both. More recently, the idea that
globular clusters may play a role in the formation of nova binaries
gained further support by the observation that the
nova rate in M87 (which has a cluster specific frequency, $S_N=14$)
appeared to be $\sim3$ times higher than that
observed for the more luminous Virgo elliptical M49, which is
characterized by a cluster specific frequency of only
$S_N=3.6$ \citep{sha00, fer03, bro06}. A weakness
of this comparison is that {\it absolute} nova rates
for different surveys are very
difficult to compare because they depend on light curve
properties, survey depth, and many other factors. The extrapolations required
to convert an observed nova rate to an absolute rate can be
large and uncertain.

To better test the dependence of nova rate on cluster specific frequency,
and thereby
the idea that a significant fraction of novae are formed in a galaxy's
globular cluster system, we initiated a survey with the
MegaCam on the CFHT 3.6 m reflector to determine the {\it relative}
nova rates in M49 and M87 using the
same telescope, instrument, and
filter, and on the same nights with the same temporal sampling \citep{sha13a}.
In a total of 4 epochs spanning 2 years of observation we discovered
a total of 29 nova candidates in each galaxy.

The spatial distributions of the nova candidates from our preliminary
analyses of M49 and M87
are shown in Figure~3. The distribution for M87 shows an
excess of nova candidates at a distance beyond 10$'$ from the
nucleus. It is possible that these transients are either intracluster
novae, or some other type of variable. If we exclude these objects,
the number of nova candidates in M87 falls to 22.
Clearly, there is no evidence from our data that the nova
rate in M87 is enhanced relative to the rate in M49, particularly
if we restrict our analysis to nova candidates that are clearly associated
with the galaxies.

\section{Populations of Novae in the Local Group}

\subsection{M31}

The number of novae in M31 with known spectroscopic class has been
increasing dramatically in recent years\footnote{See \citet{wil92} for
an introduction to the spectroscopic classification of novae.}.
At the beginning of 2010
a total of 91 spectroscopic classifications
were available for M31 novae, with 75 of these (82.4\%)
belonging to the Fe~II class \citep{sha11}. Since then,
spectroscopic classes have been determined for an additional 50 novae
bringing the total to 141.
The overall percentage of Fe~II novae has not changed significantly, with
115 novae (81.6\%) now among members of the Fe~II spectroscopic class. 
Figure~4 shows the cumulative distributions of Fe~II and He/N (+Fe~IIb)
novae in M31 plotted as a function of isophotal radius.
Although the Fe~II novae appear to be slightly more
spatially extended compared with the He/N (+Fe~IIb) novae, a KS statistic
of 0.8 indicates that this difference is not significant. A caveat to this
analysis is that the high inclination of M31's disk relative to our line
of sight makes it impossible to separate unambiguously true bulge novae from
disk novae that may be projected onto M31's bulge (or vice versa).
A better approach to
studying whether the spectroscopic class may vary with stellar population is to
compare the spectroscopic properties of novae from galaxies with differing
Hubble types.

\subsection{M33}

M33 is a relatively low mass, nearby, nearly bulgeless Local Group galaxy of
morphological type SA(s)cd \citep{dev91}.
With an estimated nova rate of only $2.5\pm1.0$~yr$^{-1}$
\citep{wil04}, little is known about the typical spectroscopic and photometric
properties of novae in this galaxy. Through the end of 2011,
a total of 36 novae had been observed in M33, including 8 with spectroscopic
classifications \citep{sha12}. Since then, 2 additional
classifications have become available from the discovery of
3 new novae. Of the 10 spectroscopic classifications, 5 are members of the
Fe~II class, with 5 members of the He/N or Fe~IIb class. The fraction
of Fe~II novae in M33 appears to be lower than that observed
in M31, but is this result significant?
If we assume the fraction of Fe~II novae in M31 is 0.82,
the probability of observing 5 or fewer Fe~II novae out of 10
with known spectroscopic class is given by:
\begin{equation}
P_{\leq 5,10} = \sum_{n=0}^5 {10! \over n! (10-n)!} \  0.82^n \, 0.18^{(10-n)} =0.021.
\end{equation}
In other words, the mix of spectroscopic nova types
in M33 differs from that of M31 at the 98\% confidence level.

\begin{figure}
\includegraphics[clip=true, scale=0.53, angle=-90]{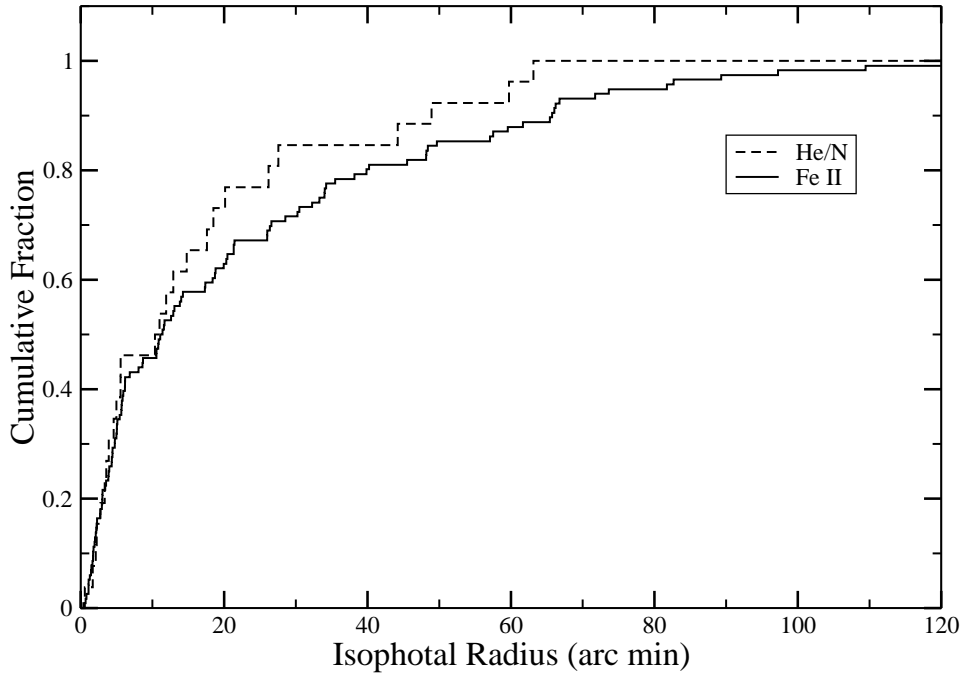}
\caption{A comparison of the cumulative distributions of Fe~II (solid line) and
He/N (+Fe~IIb) novae (broken line) in M31. A KS test shows that the observed
difference is not statistically significant.}
\end{figure}

\subsection{LMC}

Like M33, the LMC (Irr/SB(s)m) is another late-type,
low-mass Local Group galaxy with
a recent history of active star formation \citep{har09}. As described
in \citet{sha13b}, a total of 35 nova candidates have been discovered in the
LMC, 18 of which have spectroscopic data sufficient to establish
their spectroscopic classes. Of these, like in M33, 50\% belong to the
Fe~II class with the remaining 50\% belonging to either the He/N or the Fe~IIb
classes.

Of the 35 LMC novae, 29 have photometric data sufficient to estimate fade rates.
Figure~5 shows a maximum-magnitude versus rate-of-decline (MMRD) relation
for the LMC from \citet{sha13b}.
Although the scatter is significant (the magnitudes
at maximum light are often poorly determined), the most luminous novae
generally fade the fastest. Figure~6 shows the cumulative distribution
of LMC nova fade rates compared with those of M31 and the Galaxy.
In addition, as has
been noted previously \citep{del93}, novae in the LMC are generally
``faster" than novae in M31 and the Galaxy.

The fraction of recurrent novae (RNe)
in the LMC appears to be somewhat higher than 
that observed in M31 and the Galaxy (see \S 3).
Of the 38 reliable nova eruptions observed
in the LMC, 6 of them ($\sim16$\%) belong to 3 RN systems \citep{sha13b}.

\begin{figure}
\includegraphics[clip=true, scale=0.53, angle=-90]{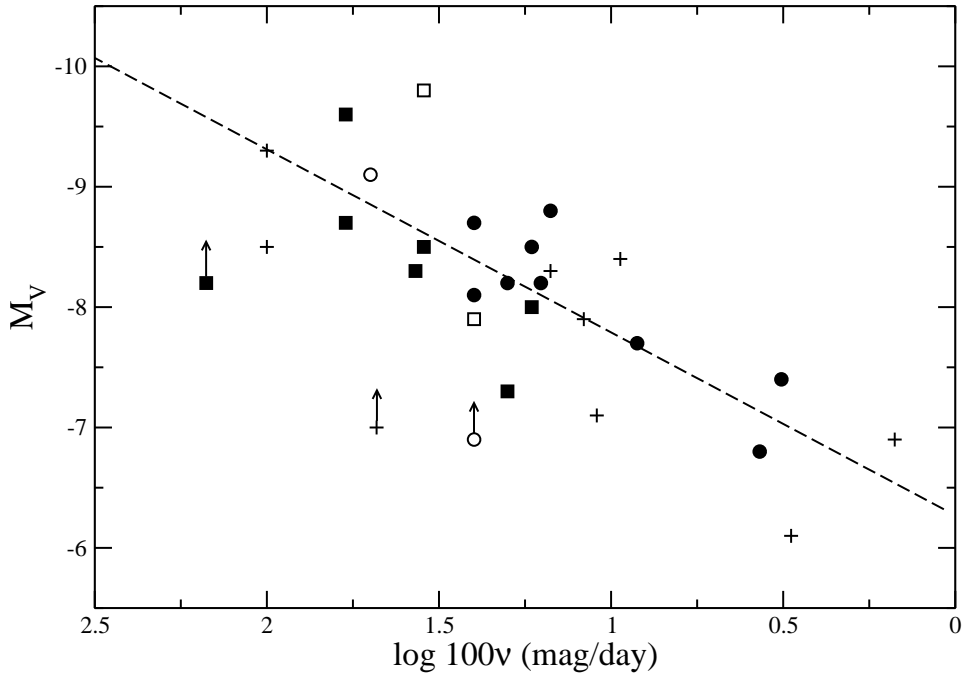}
\caption{The MMRD relation of LMC novae. Key: Fe~II novae (circles),
He/N (+Fe~IIb) novae (squares), undetermined class (+).
The dashed line is the best fitting relation. (Figure adapted
from \citet{sha13b}.)} \end{figure}

\section{Recurrent Novae in M31}

One approach to searching for RN candidates in M31 involves the
detection of the secondary star in quiescence
for those RNe that contain evolved secondaries \citep{bod09, wil13}.
This approach will identify systems that are likely to be RNe, but will miss
any short orbital period systems.
A more direct approach for identifying potential RNe involves a search
for positional coincidences among the more than 900 M31 nova candidates
discovered to
date\footnote{\tt See http://www.mpe.mpg.de/$\sim$m31novae/opt/m31/index.php}.
Given that the coordinates for novae discovered on photographic plates are not
as accurate as those for more modern discoveries, we were forced to
allow for a relatively large uncertainty in our analysis.
Our cross-correlation yielded a total of 50 nova candidates
that were spatially coincident to within 0.1$'$. Clearly,
most of these ``matches" represent chance positional coincidences. To estimate
what fraction, we performed
a Monte Carlo simulation where we randomly distributed artificial
novae with a surface density proportional to the background $R$-band
light of the galaxy, and then searched for spatial coincidences
of less than 0.1$'$ in the model distribution.
The simulation suggests that approximately
35 of the 50 spatial near coincidences observed in the real data
are expected by chance.

\begin{figure}
\includegraphics[clip=true, scale=0.53, angle=-90]{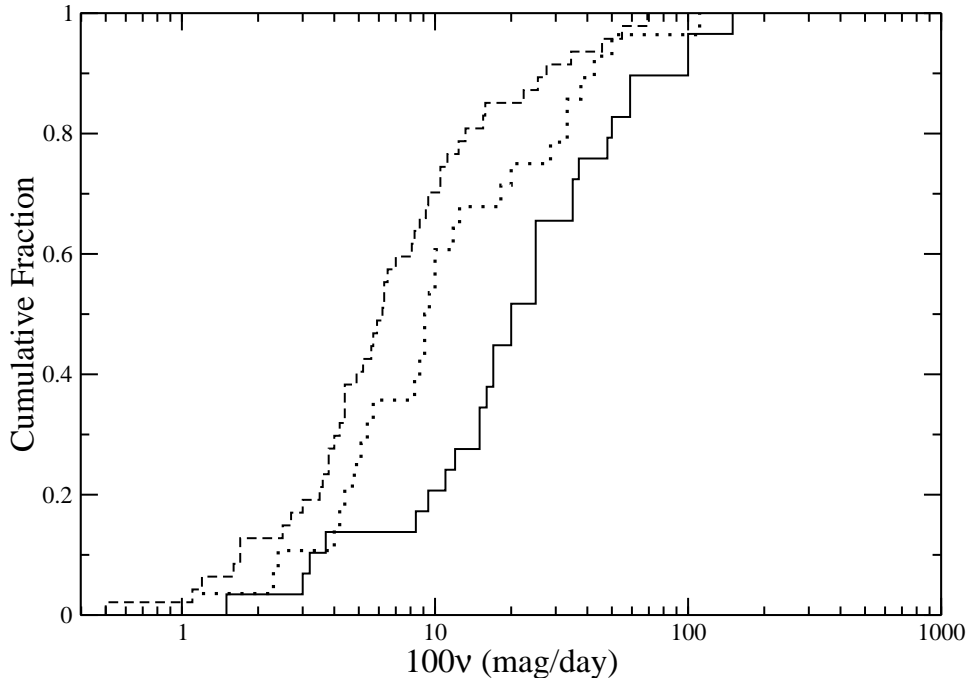}
\caption{The cumulative distribution of decline rates for LMC (solid line),
M31 (dashed line), and Galactic novae (dotted line). The LMC novae are clearly
faster on average
than those seen in M31 and the Galaxy. (Figure from \citet{sha13b},
reproduced by permission of the AAS.)}
\end{figure}

It is also possible
to narrow down the list of 50 possible RNe
by computing the probability that a specific spatial coincidence
is expected by chance. For example, a nova pair of a given separation
observed near the nucleus
(where the nova density is high) will be much more likely to be a chance
positional coincidence than would a pair with the same separation observed in
the outskirts of the galaxy.
For a given observed separation $s$, we can
compute the probability of a specific chance coincidence, $P_C$, by
considering the nova density in an isophotal annulus of area, $A$,
centered at the position of each nova. Specifically,

\begin{equation}
P_C = 1 - \rm{exp} \Big[ \sum_{i=1}^{n-1} \rm{ln} (1 - ix)\Big],
\end{equation}

\noindent
where $n$ is the number of novae in the annulus, and $x=\pi s^2/A$.

If we restrict our list of potential RNe to spatial coincidences
with separations $s\le0.1'$ and $P_C \le0.1$ we find a total of 15
RN ``strong" candidates, which agrees with the number expected
from our Monte Carlo experiments. These 15 RNe candidates represent
a total of 35 eruptions (4 of the candidates have multiple eruptions).
Of course,
the only way to be sure that any specific candidate is in fact
a RN and not a chance positional coincidence is to perform precise astrometry
on images taken during each eruption.
So far, Shafter et al. (in preparation) have been able to establish
conclusively that at least 6 of the 15 strong candidates are in fact RN.
Assuming that all 15 candidates (representing 35 eruptions)
are eventually confirmed as RNe,
we estimate that $\sim$4\% of the $\sim900$ nova eruptions observed in
M31 are due to RN, compared with $\sim$10\%
in the Galaxy and $\sim16$\% in the LMC \citep{sha13b}. Further work,
including an analysis of the observational biases in nova surveys,
will be required
before any definitive conclusions regarding the relative RN rates
in these 3 galaxies can be reached.

\acknowledgements A.W.S. acknowledges support from NSF grant AST1009566.

\bibliography{aspauthor}

\end{document}